\documentclass[12pt]{article}
\usepackage{graphicx}
\usepackage{scicite}
\usepackage{times}

\newenvironment{sciabstract}{%
\begin{quote} \bf}
{\end{quote}}

\newcounter{lastnote}

\newcommand{\ket}[1]{\mbox{$ | #1 \rangle $}}

\title{Entanglement-enabled delayed choice experiment}

\author
{Florian Kaiser,$^{1}$ Thomas Coudreau,$^{2}$ Perola Milman,$^{2,3}$\\Daniel B. Ostrowsky,$^{1}$ and S\'{e}bastien Tanzilli$^{1\ast}$\\
\\
\normalsize{$^{1}$ Laboratoire de Physique de la Mati\`{e}re Condens\'{e}e, CNRS UMR 7336,}\\
\normalsize{Universit\'{e} de Nice -- Sophia Antipolis, Parc Valrose, 06108 Nice Cedex 2, France}\\
\\
\normalsize{$^{2}$ Universit\'e Paris Diderot, Sorbonne Paris Cit\'e,}\\
\normalsize{Laboratoire Mat\'eriaux et Ph\'enom\`enes Quantiques, CNRS, UMR 7162, 75013 Paris, France}\\
\\
\normalsize{$^{3}$ Universit\'{e} Paris Sud 11, Institut de Sciences Mol\'{e}culaires d'Orsay (CNRS)}\\
\normalsize{B\^{a}t. 210, Campus d'Orsay, 91405, Orsay Cedex, France}\\
\\
\normalsize{$^\ast$To whom correspondence should be addressed; E-mail:  sebastien.tanzilli@unice.fr}
}

\date{}

\begin{document} 


\baselineskip24pt


\maketitle 


\begin{sciabstract}
Complementarity~\cite{Bohr_1984}, that is the ability of a quantum object to behave either as a particle or as a wave~\cite{Grangier_EPL_1986,Hellmut_DC_1987,Kim_DCQEraser_2000}, is one of the most intriguing features of quantum mechanics~\cite{Greenstein_Qchallenge_2006}. An exemplary Gedanken experiment, emphasizing such a measurement-dependent nature, was suggested by Wheeler using single photons~\cite{Wheeler_QTM_1984}. The subtleness of the idea lies in the fact that the output beam-splitter of a Mach-Zehnder interferometer is put in or removed after a photon has already entered the interferometer, thus performing a delayed test of the wave-particle complementary behavior~\cite{Jacques_QDC_2007}. Recently, it was proposed that using a quantum analogue of the output beam-splitter would permit carrying out this type of test after the detection of the photon and observing wave-particle superposition~\cite{Ionicioiu_ProposalQDC_2011,Schirber_WPduality_2011}.
In this paper we describe an experimental demonstration of these predictions using another extraordinary property of quantum systems, entanglement~\cite{Einstein_EPR_1935,Bell_EPR_1964,Clauser_BCHSH_1969}. We use a pair of polarization entangled photons composed of one photon whose nature (wave or particle) is tested, and of a corroborative photon that allows determining which one, or both, of these two aspects is being tested~\cite{Kaiser_Source_2011}.
This corroborative photon infers the presence or absence of the beam-splitter and until it is measured, the beam-splitter is in a superposition of these two states, making it a quantum beam-splitter. When the quantum beam-splitter is in the state \textit{present} or \textit{absent}, the interferometer reveals the wave or particle nature of the test photon, respectively. Furthermore, by manipulating the corroborative photon, we can continuously morph, via entanglement, the test photon from wave to particle behavior even \textit{after} it was detected. This result underlines the fact that a simple vision of light as a classical wave or a particle is inadequate~\cite{Ionicioiu_ProposalQDC_2011,Schirber_WPduality_2011}.
\end{sciabstract}%

While the predictions of quantum mechanics have been verified with remarkable precision, subtle questions arise when attempting to describe quantum phenomena in classical terms~\cite{Greenstein_Qchallenge_2006,Schro_Cat_1935}.
For example, a single quantum object can behave as a wave or as a particle, which is illustrated by Bohr's complementarity principle~\cite{Bohr_1984}. It states that, depending on the measurement apparatus, either wave or particle behavior is observed. This is often demonstrated by sending single photons into a Mach-Zehnder interferometer (MZI) followed by two detectors~\cite{Grangier_EPL_1986,Hellmut_DC_1987}, as represented in Fig. 1(a). If the MZI is closed, \textit{i.e.} the paths of the interferometer are recombined at the output beam-splitter (BS$_2$), the probabilities for a photon to exit at detectors $D_a$ and $D_b$ depend on of the phase difference $\theta$ between the two arms. The which-path information remains unknown, and wave-like intensity interference patterns are observed, as shown in Fig.1(b).
On the other hand, if the MZI is open, \textit{i.e.} BS$_2$ is removed, each photon's path can be known and, consequently, no interference occurs. Particle behavior is said to be observed and the detection probabilities at $D_a$ and $D_b$ are equal to $\frac{1}{2}$, independently of the value of $\theta$, as shown in Fig. 1(c). In other words, these two different configurations, \textit{i.e.} BS$_2$ present or absent, give different experimental results.
Recently it has been shown that, even when performing Wheeler's original Gedanken experiment~\cite{Wheeler_QTM_1984}, in which the configuration for BS$_2$ is chosen only after the photon has passed the entrance beam-splitter BS$_1$, Bohr's complementarity principle is still obeyed~\cite{Jacques_QDC_2007}.

In this paper we take Wheeler's experiment one step further by replacing the output beam-splitter by a quantum beam-splitter (QBS), as recently proposed theoretically~\cite{Ionicioiu_ProposalQDC_2011,Schirber_WPduality_2011}. In our realization, depicted in Fig. 2, we exploit polarization entanglement as a resource for two reasons. First, as we explain later in the text, this permits implementing the QBS. Second, this allows us to use one of the entangled photons as a test photon sent to the interferometer, and the other one as a corroborative photon. Until we detect the corroborative photon, the state of the interferometer, and therefore the wave or particle behavior of the test photon, remains unknown. By continuously modifying the type of measurement performed on the corroborative photon, we can morph the test photon from wave to particle behavior, even \textit{after} the test photon was detected.
To exclude interpretations based on either mixed states, associated with pre-existing state information, or potential communication between the two photons, the presence of entanglement is verified via violation of the Bell inequalities with a space-like separation~\cite{Bell_EPR_1964,Clauser_BCHSH_1969}.
In the following, we describe our quantum beam-splitter (QBS) apparatus. Next, we detail the action of the corroborative photon apparatus. Then, we explain the measurement procedure and present the results demonstrating that wave and particle behavior can coexist simultaneously.	

The QBS is based on the following idea. When a photon in an arbitrary polarization state enters an interferometer that is open for $\ket{H}$ (horizontally polarized) and closed for $\ket{V}$ (vertically polarized) photons, the states of the interferometer and the photon become correlated. This apparatus, shown in the right hand side of Fig. 2 and detailed in Fig. 5, therefore reveals a particle behavior for the $\ket{H}$ component of the photon state, and a wave behavior for the $\ket{V}$ component. Note that such an experiment has been realized using single photons prepared in a coherent superposition of $\ket{H}$ and $\ket{V}$~\cite{Tang_BorhPrinciple_2012}. However, we take this idea to a new level, achieving genuine quantum behavior for the output beam-splitter by exploiting an intrinsically quantum resource, entanglement. This allows entangling the quantum beam-splitter and test photon system with the corroborative photon. Thus, measurement of the corroborative photon enables projecting the test photon/QBS system into an arbitrary coherent wave-particle superposition, which is a purely quantum object.
In other words, our QBS is measured by another quantum object, which projects it into a particular superposition of present and absent states. 
More precisely, we use as a test photon one of the photons from the maximally polarization entangled Bell state, $\ket{\Phi^+} = \frac{1}{\sqrt{2}} \left( \ket V_c \ket V_a + \ket H_c \ket H_a\right)$, produced at the wavelength of 1560\,nm using the source described in~\cite{Kaiser_Source_2011}. Here, following the notation of Fig. 2, $c$ and $a$ denote the spatial modes of the photons. This ensures maximum randomness of the input polarization state of the test photon ($a$), which enters a Mach-Zehnder interferometer with a QBS for the output beam-splitter.

The actual QBS device is made up of two components. The first is a polarization dependent beam-splitter (PDBS) that shows close to 100\% transmission for horizontally polarized photons and provides ordinary 50/50 splitting ratio for the vertically polarized photons. The PDBS is realized using a combination of standard bulk optical components (see Fig. 5 in the supplementary information). The whole state after the PDBS reads

\begin{equation}
	\ket{\Psi} = \frac{1}{2} \left( c^{\dagger}_H \, \left(  a^{\dagger}_H + i\,e^{i\,\theta} \, b^{\dagger }_H \right) + \frac{1}{\sqrt{2}}c^{\dagger}_V \,  \left( b^{\dagger}_V (i+i\,e^{i\,\theta}) + a^{\dagger}_V (1-\,e^{i\,\theta})\right) \right)\ket{vac}.
\end{equation}
\noindent
Here, using the notation of Fig. 2, $c^{\dagger}_H$ and $c^{\dagger}_V$ represent creation operators for horizontally and vertically polarized photons, respectively, propagating towards the corroborative photon apparatus, and $\theta$ is an adjustable phase shift. Similarly, $a^{\dagger}$ and $b^{\dagger}$ symbolize creation operators for test photons propagating toward PBS$_1$ and PBS$_2$, respectively. Moreover, $\ket{vac}$ represents the vacuum state.
At this point (see Eq. 1), each polarization state of the test photon is associated with one of the two complementary types of behaviors, wave and particle.\\
The second stage consists of polarizing beam-splitters (PBS$_1$ and PBS$_2$) oriented at 45$^{\circ}$ to the $\{H, V\}$ basis, that erase all polarization information that existed at the PDBS output~\cite{Kim_DCQEraser_2000,Herzog_Complement_1995}. The state of Eq.~1 is now described by the following equation

\begin{equation}
 \ket{\Psi} = \frac{1}{\sqrt{2}} \left( c^{\dagger}_H [particle]^{\dagger} + c^{\dagger}_V [wave]^{\dagger} \right)\ket{vac},
\end{equation}
with $$[particle]^{\dagger} = \frac{1}{2}\left( a^{\dagger}_H + a^{'\dagger}_V + i\,e^{i\,\theta} (b^{\dagger}_H + a^{'\dagger}_V)  \right),$$
\noindent and $$[wave]^{\dagger} = \frac{1}{2\,\sqrt{2}}\left( (1-e^{i\,\theta}) (-a^{\dagger}_H + a^{'\dagger}_V) + (i\,e^{i\,\theta}+i) (-b^{\dagger}_H + b^{'\dagger}_V)\right),$$
\noindent
where the creation operators $a^{\dagger}$, $a^{'\dagger}$, $b^{\dagger}$ and $b^{'\dagger}$ denote photons propagating toward detectors $D_a$, $D_{a'}$, $D_b$ and $D_{b'}$, respectively.
Consequently, the only way of knowing if wave or particle behavior was observed is by examining the corroborative photon.

We will now describe the corroborative photon measurement apparatus, as shown on the left-hand side of Fig. 2.
It consists of two stages. The first is an electro-optic phase modulator (EOM) that allows rotating the polarization state of the corroborative photon by an angle $\alpha$. 
From Eq. 2, we now have
\begin{equation}
 \ket{\Psi} = \frac{1}{\sqrt{2}} \left( c^{\dagger}_H \left( \cos \alpha [particle] - \sin \alpha [wave] \right) + c^{\dagger}_V \left( \cos \alpha [wave] + \sin \alpha [particle] \right)\right)\ket{vac}.
\end{equation}
After passing PBS$_3$, that is oriented on the $\{H,V\}$ axis, the corroborative photon is transmitted $(\ket H)$ or reflected $(\ket V)$. This projects the test photon into a state defined by the terms in the parentheses of Eq. 3. Therefore, the firing of detector $D_H$ indicates that the test photon is in the state $\cos \alpha [wave] - \sin \alpha [particle]$, while the firing of $D_V$ that it is in the state $\cos \alpha [particle] + \sin \alpha [wave]$. It can be seen that by choosing $0<\alpha< 90^{\circ}$, a continuous morphing between wave and particle behavior is obtained.
The expected intensity correlations, given by the coincidence count probability between detectors $D_H$ (corroborative) and $[D_a \oplus D_{a'}]$ (test), where $\oplus$ denotes an exclusive OR (XOR) gate, are
\begin{eqnarray}
 I_a (\theta,\alpha) =\cos^2 \frac{\theta}{2} \, \sin^2 \alpha + \frac{1}{2} \cos^2 \alpha.
\end{eqnarray}
Note that the correlations between detectors $D_V$ and $[D_b \oplus D_{b'}]$ follow the same function. On the contrary the complementary intensity correlations, \textit{i.e.} between detectors $D_V$ and $[D_a \oplus D_{a'}]$ or between $D_H$ and $[D_b \oplus D_{b'}]$, are given by $1-I_a(\theta,\alpha)$. The use of XOR gates permits counting the photons from both outputs of each quantum eraser (PBS$_{1}$ or PBS$_2$), and reaching an average coincidence rate of 70/s for each of them. Note that Eq. 4 does not depend on the relative detection times of the two photons. In the experiment reported here, the detection of the corroborative photon is delayed until \textit{after} the detection of the test photon. This is ensured by inserting an extra 5\,m length of optical fiber in the path of the corroborative photon ($c$).
In this case, for all the four correlation functions mentioned above, the configuration of the interferometer remains undetermined even after the test photon has been detected. In other words, there is no information available yet from the corroborative photon that could influence the behavior of the test photon.
Furthermore, a space-time analysis shows that no classical communication can be established between the photon detection events, as they are space-like separated, as shown in Fig. 3.

We now measure the correlations between detectors $D_H$ and $[D_a \oplus D_{a'}]$ via counting coincidence events on the corresponding single photon detectors (InGaAs avalanche photo diodes). As shown in Fig. 4(a, b), the experimentally measured and theoretically expected results are in near perfect agreement. For the angle $\alpha=0^{\circ}$, $I_a (\theta,0)$ is independent of the phase $\theta$ as predicted for particle-like behavior. Setting $\alpha=90^{\circ}$ results in sinusoidal intensity oscillations as a function of $\theta$, which corresponds to wave-like behavior. For $0^{\circ} < \alpha < 90^{\circ}$, a continuous transition from wave to particle behavior is observed, expressed by the continually reducing fringe visibility.

To prove the existence of a coherent quantum superposition of wave and particle behavior of the test photon created by the detection of the corroborative photon, the presence of entanglement needs to be verified~\cite{Einstein_EPR_1935,Bell_EPR_1964,Clauser_BCHSH_1969,Kaiser_Source_2011}. Note that several recent realizations ignored this and therefore the presence of a QBS has not been proven unambiguously~\cite{Roy_NMR_2012,Auccaise_Qcomple_2012}. In our realization, entanglement is proven by performing the same experiment as before, but using the complementary analysis basis, namely the diagonal basis $\{D,A\}$. Now, the initial quantum state is rotated by 45$^{\circ}$, \textit{i.e.} $\frac{1}{\sqrt{2}} \, \left(  c^{\dagger}_V a^{\dagger}_V + c^{\dagger}_H a^{\dagger}_H \right) \rightarrow \frac{1}{\sqrt{2}} \, \left(  c^{\dagger}_D a^{\dagger}_D + c^{\dagger}_A a^{\dagger}_A \right)$, where $D$ and $A$ symbolize diagonally and anti-diagonally polarized photon contributions, respectively. In this configuration, every single photon is unpredictably subjected to a closed or open Mach-Zehnder configuration by the PDBS. In this case, as opposed to the experiment in the $\{H,V\}$ basis, if a statistical mixture was analysed instead of an entangled state, no correlations should be observed when measuring $I_a (\theta,\alpha)$. However, the strong correlations shown in Fig. 4(b) exclude a statistical mixture and are in good agreement with the theoretical expectation represented in Fig. 4(d). This underlines that wave and particle behavior coexist simultaneously for the entire range $0^{\circ} < \alpha < 90^{\circ}$ in the $\{H,\,V\}$ basis, and for $-45^{\circ} < \alpha < 45^{\circ}$ in the $\{D,\,A\}$ basis. The quality of the entangled state is measured via the Bell parameter $S$, which is deduced from the phase oscillation visibilities at $\alpha=90^{\circ}$ in the $\{H,\,V\}$ basis, and $\alpha=45^{\circ}$ in the $\{D,\,A\}$ basis. We obtain $S=2.77\pm0.07$, which is very close to the optimal value of $2\sqrt{2}$ attained with maximally entangled states, and 11 standard deviations above the classical/quantum boundary $S = 2$~\cite{Einstein_EPR_1935,Bell_EPR_1964,Clauser_BCHSH_1969,Kaiser_Source_2011}.

We note that the 'detection loophole' remains open in our experiment, since some of the initial entangled photons are lost during their propagation in the fiber or bulk channels, or are not detected by the single photon detectors that show non-unit quantum detection efficiencies~\cite{Pearle_HV_1970,Clauser_LTheories_1974}. One has therefore to assume that non detected photons behave identically to the detected ones, as suggested by the fair-sampling theorem~\cite{Clauser_BellTheo_1978}.

In conclusion, we have carried out a quantum delayed choice experiment, enabled by polarization entangled photons and the associated property of non-locality. We employed a Mach-Zehnder interferometer where the output beam-splitter has been replaced by its quantum analogue, \textit{i.e.} a beam-splitter in a coherent superposition of being present and absent. In this configuration, we observed that the single photons under test can indeed behave as waves and particles in the same experiment. We experimentally excluded interpretations based on local hidden variables and/or information exchange between the photon and the quantum beam-splitter. This was done by revealing the state of the quantum beam-splitter only after the detection of the corroborative photon. We have, therefore, demonstrated delayed interference between wave and particle behavior, which underlines the subtleness of Bohr's complementarity principle.

We note that, parallel to this work, Peruzzo \textit{et al.} realized another version of a quantum delayed choice experiment based on entangled photons~\cite{Peruzzo_QDC_2012}.

The authors thank L. A. Ngah for help on data acquisition, and O. Alibart for fruitful discussions. Financial support from the CNRS, l'Universit\'e de Nice - Sophia Antipolis, l'Agence Nationale de la Recherche for the 'e-QUANET' project (grant ANR-09-BLAN-0333-01), the European ICT-2009.8.0 FET open program for the 'QUANTIP' project (grant 244026), le Minist\`{e}re de l'Enseignement Sup\'{e}rieur et de la Recherche (MESR), la Fondation iXCore pour la Recherche, and le Conseil R\'{e}gional PACA for the 'QUANET' project.

\newpage

\includegraphics[width=0.9\columnwidth]{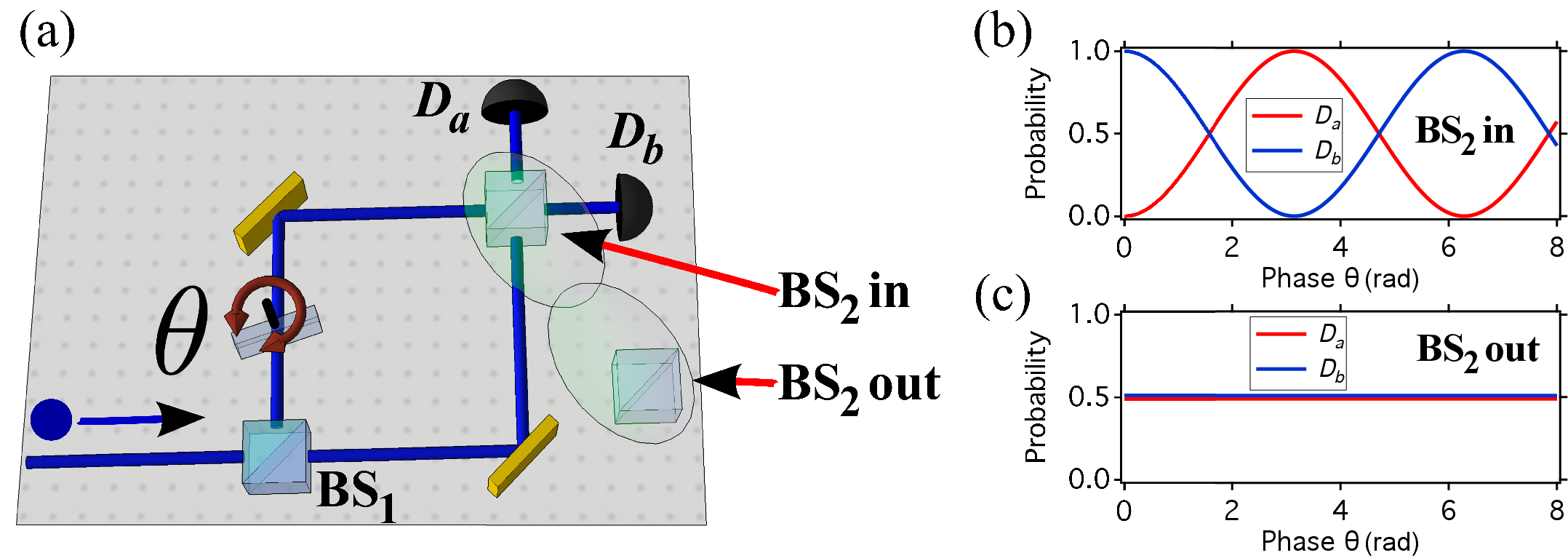}\\
Fig. 1: (a) - Explanation of Wheeler's 'Gedanken' experiment using a Mach-Zehnder interferometer. The device consists of two beam-splitters, BS$_1$ and BS$_2$, and a glass plate introducing a phase shift $\theta$. When a single photon enters the device at BS$_1$, it is split into a coherent, \textit{i.e.} quantum, superstition of travelling along the transmission and reflection paths. A second beam-splitter (BS$_2$) can be put in or out at the will of the user to generate an open (closed) Mach-Zehnder interferometer with detectors $D_a$ and $D_b$ at its output. (b) -  Simulated photon detection probabilities at detectors $D_a$ respectively $D_b$ as a function of the phase $\theta$. The sinusoidal oscillations are related to unknown path information, and therefore to single photon interference, which is a wave-like phenomenon. (c) - Detection probabilities without BS$_2$. No interference is observed, as the single photon paths can be traced back, which is the signature of particle behavior.
\vspace{2cm}\\

\includegraphics[width=0.9\columnwidth]{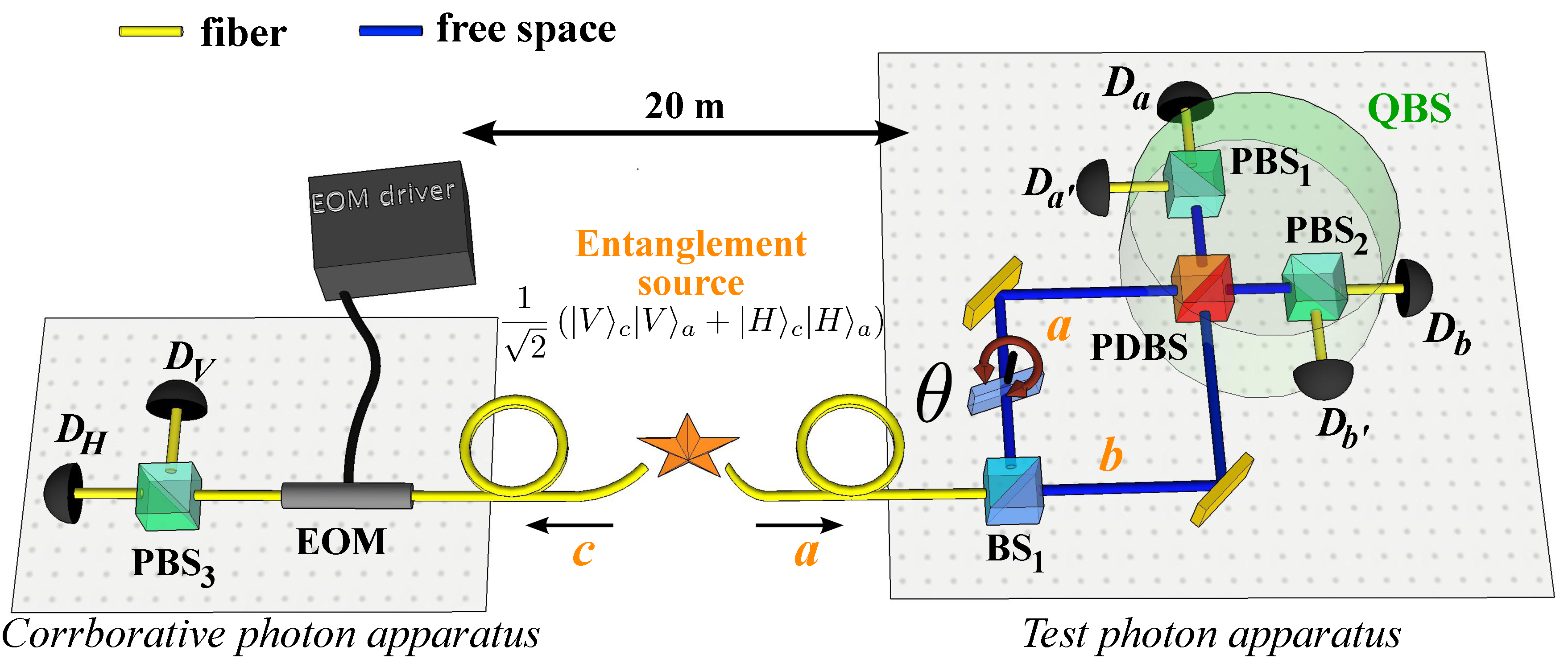}\\
Fig. 2: Experimental setup. A source of polarization entangled photons ($\lambda$\,=\,1560\,nm, see (13) for more details) sends, through a single mode optical fiber, one photon ($a$) to a quantum beam-splitter (QBS) apparatus, being an open (resp. closed) Mach-Zehnder interferometer for horizontally (resp. vertically) polarized photons. This is realized using a polarization dependent beam-splitter (PDBS), showing a 50/50 splitting ratio for vertical, and 100/0 ratio for the horizontal, components, respectively. The polarization state information on this photon is erased via projection on polarizing beam-splitters (PBS$_{1,2}$) oriented at 45$^{\circ}$ compared to the PDBS basis. The second photon ($c$) of the entangled state is sent to another laboratory 20\,m away, and used as a 'corroborative' photon and allows determining whether wave or particle-like behavior was observed. The space-like distant QBS corroborative photon apparatus is used for rotating the polarization state of this photon using an electro-optical modulator (EOM). The photon then goes through PBS$_3$, followed by two detectors ($D_H$ and $D_V$)), one for each projected polarization state.
The paths are denoted as follows. Photons going to the QBS corroborative apparatus are said to propagate along path $c$. A photon going toward the QBS apparatus travels along path $a$. In the QBS apparatus, the upper path is denoted by $a$, while the lower path are denoted by $b$. Finally, after PBS$_{1,2}$, the paths toward detectors $D_a$, $D_{a'}$, $D_b$ and $D_{b'}$ are denoted by $a$, $a'$, $b$ and $b'$, respectively.
\\

\includegraphics[width=0.7\columnwidth]{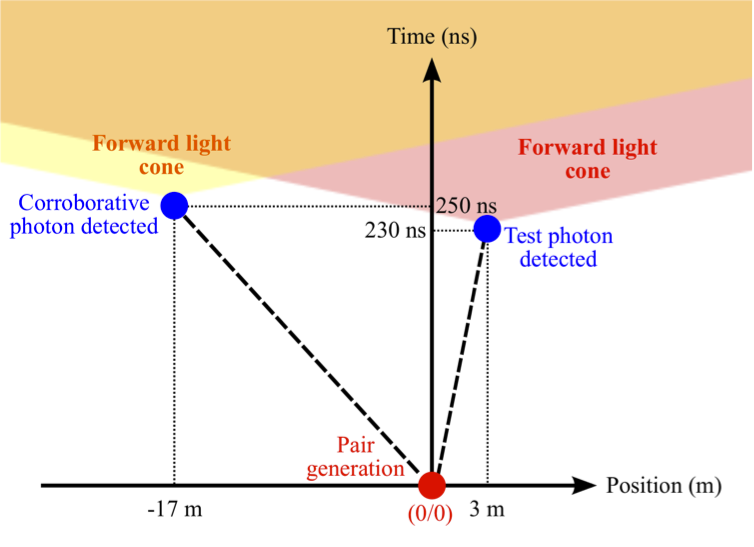}\\
Fig. 3: Space-time diagram of the experimental apparatus. The paired photons are said to be generated and separated at the origin (0/0). The test photon travels about 50\,m in an optical fiber before entering the QBS apparatus, that is located in the same laboratory as the entangled photon pair source. The corroborative photon is sent through a 55\,m fiber to another laboratory. The corroborative and test photon apparatus are physically separated by 20\,m. Note that the corroborative photon is measured 20\,ns after the test photon was detected, thus revealing the Mach-Zehnder interferometer configuration in a delayed fashion. The forward light cones from both photon detection events do not overlap, demonstrating that space-like separation is achieved. In other words, no causal connection between these events can be established. \vspace{2cm}\\

\begin{tabular}{cc}
\vspace{-2cm}\\
\includegraphics[width=0.5\columnwidth]{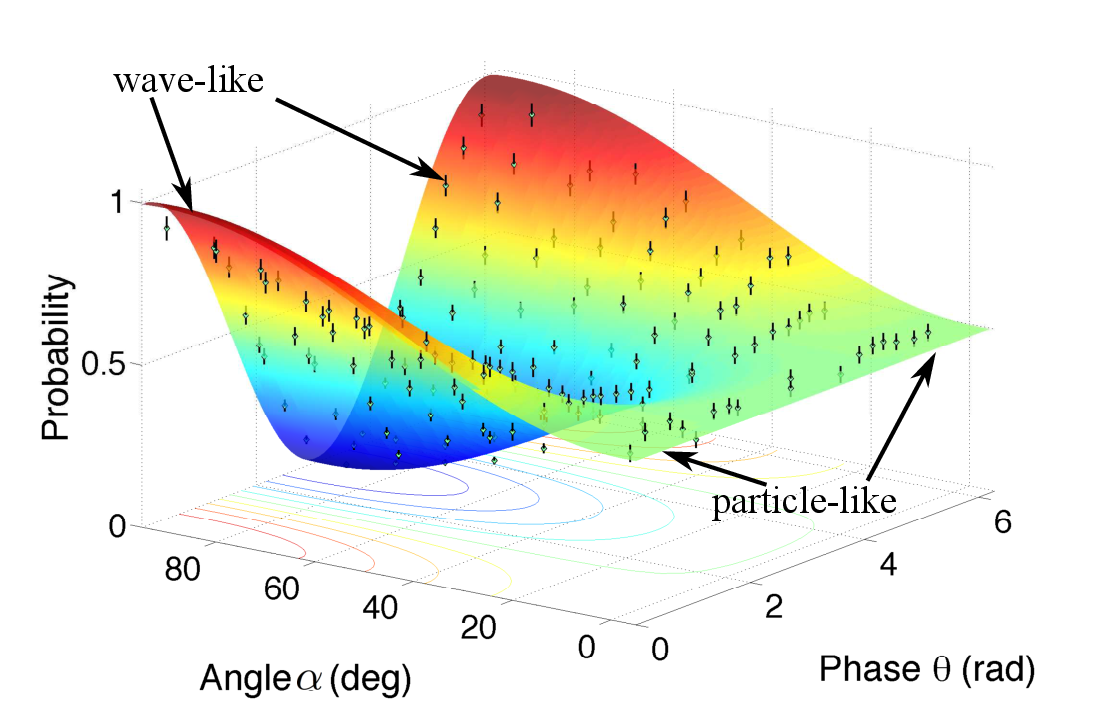} & \includegraphics[width=0.5\columnwidth]{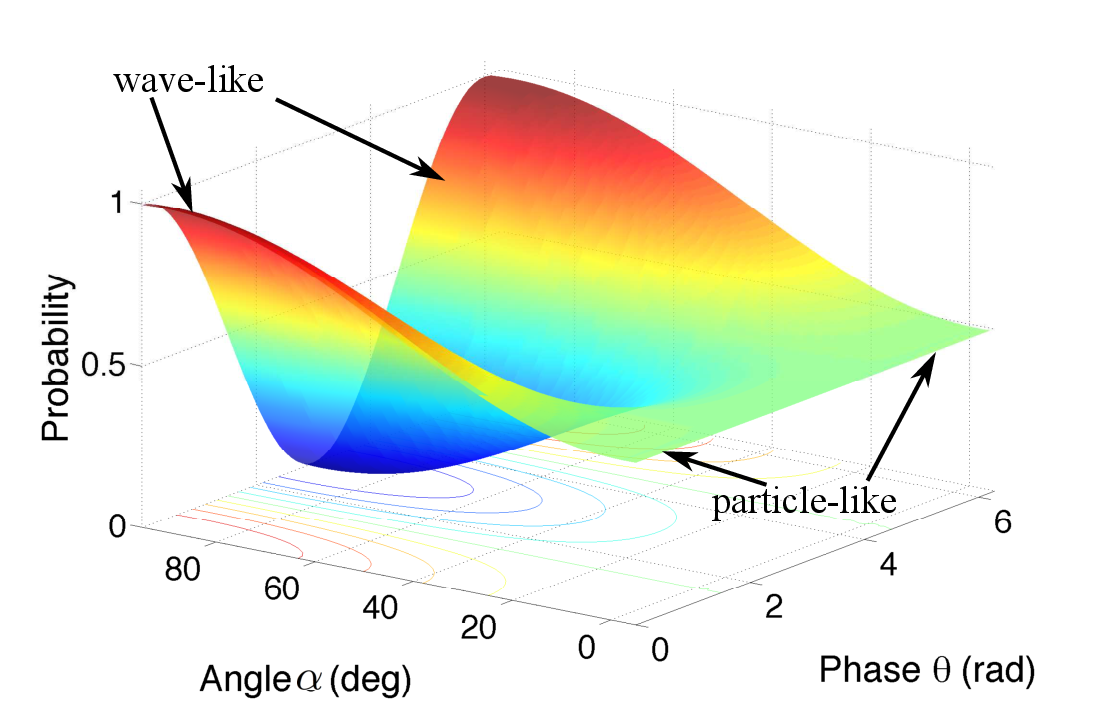}\\
(a) & (b)\\
\includegraphics[width=0.5\columnwidth]{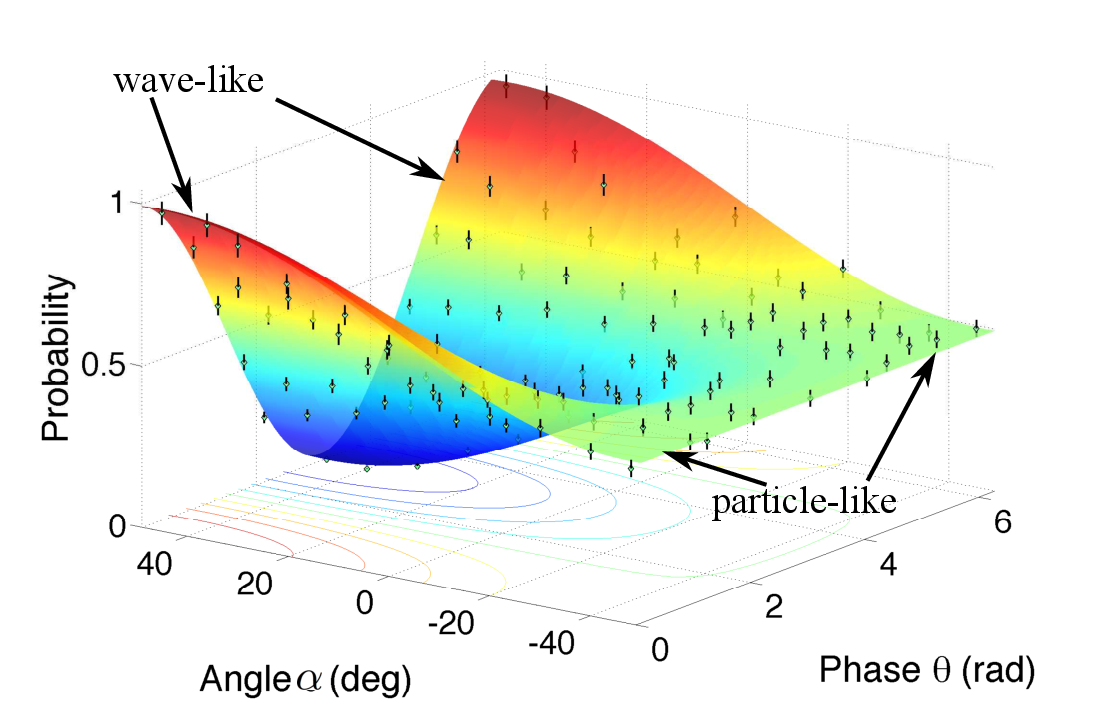} & \includegraphics[width=0.5\columnwidth]{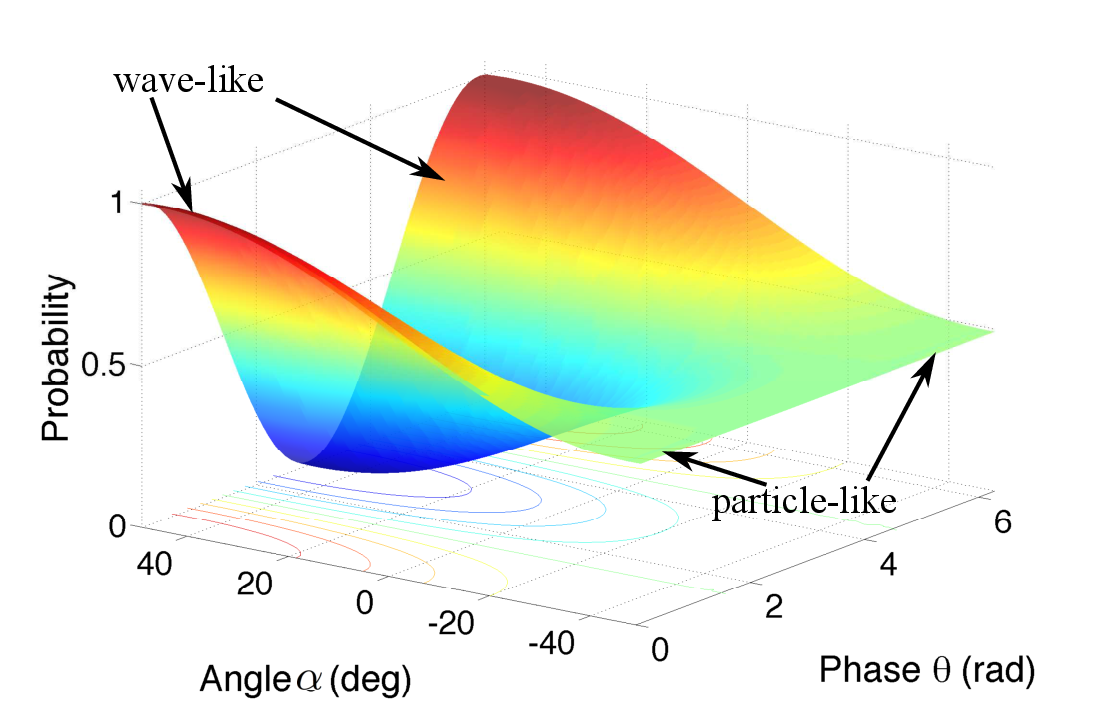}\\
(c) & (d)\\
\end{tabular}

\noindent
Fig. 4: Experimental results for the quantum delayed choice experiment. Plotted are the intensity correlations, $I_a(\theta,\alpha)$ as defined by Eq.~4, expressed as the probability of a coincidence event between detectors $D_H$ and ($D_a \oplus D_{a'}$) as a function of $\alpha$ and $\theta$.
Dots and associated vertical lines represent experimental data points and their corresponding standard deviations.
Wave-particle morphing behavior is observed for the natural $\{H,V\}$ basis (a), as well as for the complementary $\{D,A\}$ basis (c). The colored surfaces in these graphs represent the best fits to the experimental data using Eq. 4. Note that the result obtained for the $\{D,A\}$ basis is essential since it represents the signature of the entangled state, proving the correct implementation of the desired quantum beam-splitting effect.
Figures (b) and (d) represent the simulation of the results expected by quantum theory for the $\{H,V\}$ and $\{D,A\}$ basis, respectively. Note that similar results have been obtained for the three other intensity correlation functions as defined in the text (not represented). Experimentally, we obtain average coincidence rates of 350 events/5\,s. The noise contribution, on the order of 3 events/5\,s, has not been subtracted. Note that the same experimental results would be obtained when inverting the timing order of the measurements of the test and corroborative photons, similar to the behavior reported in~\cite{Ma_QCDSwapp_2012}.

\newpage

\section*{Supplementary information}
Note that the above introduced polarization dependent beam-splitter (PDBS) has been mimicked using a set of standard bulk optical components toward achieving high quality experimental results. The schematic is shown in Fig. 5.\\

\includegraphics[width=0.7\columnwidth]{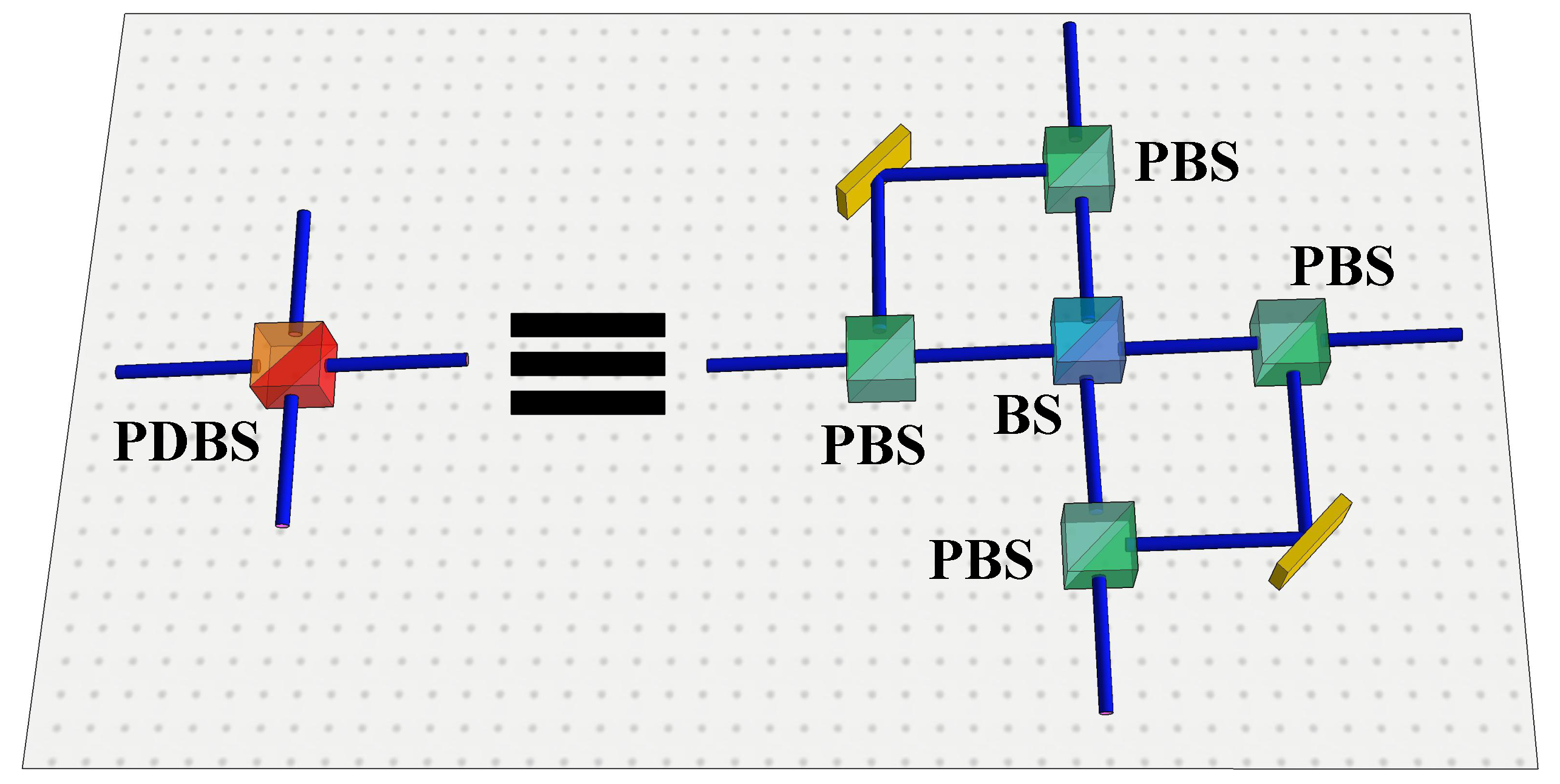}\\
Fig. 5: In this realization the polarization dependent beam-splitters (PDBS), yielding the desired 100/0 transmission/reflection ratio for $\ket H$ and 50/50 transmission/reflection ratio for $\ket V$ was built using the bulk configuration shown. We used four polarizing beam-splitters (PBS) oriented in the $\{H,V \}$ basis. While $\ket H$ photons are reflected on each PBS and bypass the beam-splitter (BS), $\ket V$ photons are transmitted to an ordinary 50/50 BS. Commercially available PDBS devices show seriously reduced performance and would have significantly reduced the measured visibilities.

\end{document}